\documentclass[aps,pra,amssymb,twocolumn,showpacs,supersriptaddress,groupeaddress]{revtex4-1}
\usepackage{amssymb}
\usepackage{amsmath}
\usepackage{graphicx}
\usepackage{float}
\usepackage{dcolumn}
\usepackage{bm}
\usepackage{natbib,hyperref}
\usepackage{hyperref}

\usepackage{color}
\hypersetup{colorlinks=true, 
    linkcolor=red,          
    citecolor=magenta,        
    filecolor=gree,      
    urlcolor=blue           
}

\usepackage[utf8]{inputenc}
\usepackage{graphicx}
\usepackage{dcolumn}
\usepackage{bm}
\usepackage{xcolor}
\usepackage{amsmath}
\usepackage{amssymb}
\usepackage{braket}
\usepackage{float}

\begin{document}

\preprint{APS/123-QED}

\title{Dissipative entanglement generation  between  two driven qubits in circuit quantum electrodynamics}

\author{Sebastián Luciano Gallardo$^{1}$, Daniel Domínguez$^{1}$ and María José Sánchez$^{1,2}$}
\affiliation{$^{1}$Centro At\'omico Bariloche and Instituto Balseiro (Universidad Nacional de Cuyo), 8400 San Carlos de Bariloche, R\'io Negro, Argentina.\\
$^{2}$Instituto de Nanociencia y Nanotecnolog\'{\i}a (INN),CONICET-CNEA, Argentina.}

\date{\today}

\begin{abstract}

An entangled state generation protocol for a system of two qubits driven with an ac signal and coupled through a resonator is introduced.  
We explain the mechanism of entanglement generation in terms of an interplay between unitary  Landau-Zener-St\"uckelberg (LZS)  transitions  induced for appropriate  amplitudes and frequencies  of the applied ac signal and  dissipative processes dominated by photon loss. 
In this way, we found that the steady state of the system can be tuned to be arbitrarily close to a Bell state, which  is independent  of the initial state.
Effective two-qubit Hamiltonians that reproduce the resonance patterns associated with LZS transitions are derived.

\end{abstract}

\maketitle

\section{Introduction}




The generation and stabilization of entangled states is of fundamental importance for quantum information applications. 
In the last two decades, several proposals have explored strategies based on the use of environmental noise to obtain and stabilize  steady state entanglement \cite{kraus_2008,verstraete_2009,tacchino_2018}.
 
Most of these schemes use an external  driving
field as a tool, with examples including adiabatic passage protocols \cite{Kral_2007} - extensively employed to generate quantum state transfer \cite{Maeda_2006, Zhou_2017},  weak resonant drivings - which enable entanglement stabilization based on tailoring the relaxation rates in order to generate a  highly entangled steady state \cite{shankar_2013,leghtas_2013,kimchi_2016, Quintana_2013, champagne_2018}, or a frequency-modulated signal   \cite{li_2020} - used to achieve an accelerated formation of dissipative entangled steady states.

These  protocols have been tested in several systems such as  atomic ensembles \cite{krauter_2011},  trapped ions \cite{barreiro_2011,lin_2013, kienzler_2015}, Rydberg atoms \cite{li_2020, Maeda_2006} and superconducting qubits \cite{dicarlo_2010, shankar_2013,leghtas_2013, kimchi_2016, Quintana_2013,champagne_2018}, to mention a few.

Recently  a mechanism relying on the  amplitude-modulation of an ac signal was  proposed to generate  steady-state entanglement in  a system of two coupled qubits driven by a large amplitude (non resonant) periodic signal  and interacting with  a thermal bath \cite{gramajo_2018, gramajo_2021}.

Nowadays, circuit quantum electrodynamics (cQED)  \cite{blais_2004,wallraff_2004, xiang_2013,Blais_2020,Blais_2021} has been established as one of the leading architectures for
studying quantum computation and quantum simulation, where superconducting qubits are connected to a transmission line resonator \cite{koch_2007,Gu_2017, Bonifacio_2020}.
Many important experimental advances have been achieved
in this regard, including  the observation of Jaynes-Cummings ladder \cite{fink_2009} and  long-lived qubit-resonator states \cite{paik_2011}, entanglement of distant qubits, realization of one and two qubit gates and non-demolition readout operations \cite{esteve_2012,dicarlo_2009, dicarlo_2010, stern_2014,  johnson_2010, walter_2017, van_loo_2013, eichler_2012, didier_2015,campagne-ibarcq_2018}.

In this work we propose a protocol to generate and stabilize maximally entangled states (in particular, Bell states) in a system of two qubits driven  with an ac signal,  which are indirectly coupled  via  a common resonator. Although this driving protocol has been  implemented in studies of  Landau-Zener-St\"uckelberg interferometry  \cite{Shevchenko_2010, Oliver_2009, Ferron_2012, Ferron_2016, gramajo_2019, Bonifacio_2020}
and entanglement generation \cite{gramajo_2018,gramajo_2021} with superconducting qubits, we are not aware of previous  proposals  employing ac driven qubits to control entanglement in cQED architectures. Our approach is rather general and not restricted to the usual weak resonant driving, going  beyond the standard  dispersive regime used to couple the resonator for readout \cite{Blais_2020}. 
As it is customary in  cQED architectures, we will  assume that the resonator acts as a filter of noise for the qubits \cite{Bronn_2015,Blais_2020}, protecting them from spontaneous losses to the environment.  With this in mind, we will model  the  environment as a thermal bath coupled to the system mainly through the resonator. 

Through an interplay between  driving and dissipation, we show that an  unique stationary maximally entangled  (Bell) state can be obtained  regardless  the initial state of the system, provided the qubits are driven with the appropriate amplitude and frequency. Moreover, the obtained  Bell state is protected from environmental effects for as long as the driving is applied.


 The paper is organized as follows: in Sec.\ref{sec:sys_overview} we do an overview of the system and present the model Hamiltonian. In Sec.\ref{sec:unitary} we  solve the unitary driven  dynamics of the system and analize  the  LZS resonance patterns of the  two  relevant transitions involved
 in the generation of maximally entangled steady states, once coupling to environment is included in Sec.\ref{sec:bell}.
Additionally two-qubit Hamiltonians that reproduce the structure of these resonances are also derived in  Sec.\ref{sec:unitary}.
Conclusions and perspectives are given in Sec.\ref{sec:concl}.

\section{System overview\label{sec:sys_overview}}

\begin{figure}
    \centering
    \includegraphics[width=\linewidth]{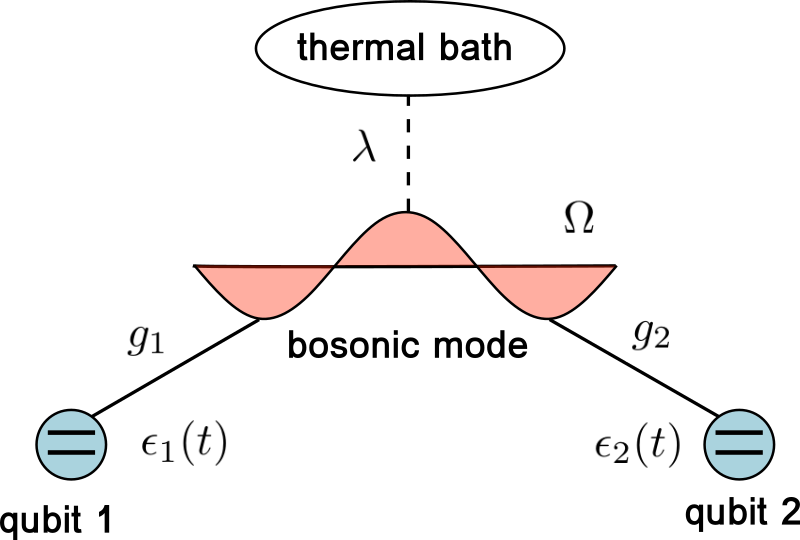}
    \caption{Schematic representation of the system under study. The qubit $i\in \{1,2\}$ is  periodically driven through $\epsilon_i(t)$  and  coupled with strength $g_i$ to a mode of frequency $\Omega$ of a resonator. The resonator  is weakly coupled to a thermal bath with strength $\lambda$ via the bosonic operator $a+a^\dagger$.}
    \label{fig:systemscheme}
\end{figure}

We  study a system composed of two qubits coupled to a bosonic mode within a resonator, which is itself weakly coupled to a thermal bath with temperature $T_\text{b}$, as   is shown schematically in Fig. \ref{fig:systemscheme}.

The Hamiltonian of the qubit $i\in \{1,2\}$, including the coupling term to the resonator is given by

\begin{equation}\label{eq:Hi}
    H_i(t) = \frac{\epsilon_i(t)}{2}\sigma_z^{(i)} + g_i (a+a^\dagger)\sigma_x^{(i)},
\end{equation}
where  $\sigma_j^{(i)}$ are the Pauli matrices acting on the qubit $i$, and  $a^\dagger$ ($a$) is the creation (destruction) operator of the bosonic mode of the resonator. The qubit $i$'s  transition frequency or detuning is $\epsilon_i$ , which
   can be controlled externally as a function of time.  The coupling of the qubit $i$ to the resonator is of strength $g_i$ and  we suppose that the associated operator is  transversal to the qubit $i$'s detuning operator (under this assumption, one can always rotate the qubit basis such that the coupling to the resonator is through $\sigma_x^{(i)}$). A possible qubit $i$ energy gap induced by a term  in Eq.(\ref{eq:Hi}) transversal to  $\sigma_z^{(i)} $  was neglected under the assumption that it is much smaller than the corresponding $g_i$'s  -  which is  rather justified for several  superconducting qubit systems \cite{Oliver_2005}. 
The full Hamiltonian for the cQED architecture  is given by
\begin{align}
    &H(t) = H_\text{s}(t) + H_\text{b} + H_\text{sb},\label{eq:H_3parts}\\ 
    &H_\text{s}(t) = \Omega a^\dagger a + \sum_{i=1}^2 H_i(t) \label{eq:H0t},
\end{align}
where  $\Omega$ is the resonator mode frequency. The term $H_\text{b}$ in Eq.(\ref{eq:H_3parts}) represents the  bath Hamiltonian, modelled as a continuum of harmonic oscillators in thermal equilibrium at temperature $T_\text{b}$, with ohmic spectral density $\mathcal{J}(\omega) = \kappa\omega$, where $\kappa$ is a constant.
The term $H_\text{sb}$ stands for the interaction between the system and the thermal bath, which in this work we suppose is through the  operator $(a+a^\dagger)$ and of strength $\lambda$. The explicit forms of $H_\text{b}$ and $H_\text{sb}$ are  given in Appendix \ref{app:env}.
 
The driving required for the Bell state generation protocol depends on the relative sign of the couplings $g_i$. We suppose that the couplings are similar in magnitude but their relative sign could be either equal or opposite, corresponding  to couplings to even or odd modes of the resonator, respectively. 
In the following without loss of generality we will consider couplings with the same sign  and  the drivings in detuning chosen as
\begin{equation}
    \epsilon(t) \equiv \epsilon_1(t) = \epsilon_2(t) = A\cos(\omega t)\label{eq:driving},
\end{equation}
with $A$ the amplitude and $\omega$ the frequency of the driving. For the  case of opposite coupling signs, the driving should be chosen to be $\epsilon_1(t)=-\epsilon_2(t)$. It can be shown that both cases are related by a local unitary transformation $H\rightarrow \sigma_y^{(2)}H\sigma_y^{(2)}$ which keeps the entanglement generation dynamics invariant. 

As we will discuss in detail in Sec.\ref{sec:bell}, to stimulate the LZS resonances necessary for this  Bell state generation protocol  we require that $0 <  |\delta_g | \ll g_1 g_2/\Omega$, with $\delta_g \equiv g_1-g_2$.
As  the relevant involved transitions  occur in a timescale $\delta_g^{-1}$,  the smaller $\delta_g$, the longer it will take  to reach the stationary Bell state. 




To solve numerically the dynamics  in the purely unitary case (considering only $H_\text{s}(t)$ in Eq.(\ref{eq:H_3parts})), we  diagonalize the evolution operator over a period of the driving using a 4$^\text{th}$ order Trotter-Suzuki expansion. In this way we obtain the Floquet states and  the associated quasienergies  \cite{Shirley_1965}.
To study the open dynamics we evolved the system's density operator using the Floquet-Born-Markov (FBM) master equation \cite{kohler_1998,Blattmann_2015, Ferron_2016} within a moderate Rotating Wave Approximation (RWA), as is detailed in Appendix \ref{app:fbm}. 
For the numerical simulations we truncate the Hilbert space to  a finite number of photon levels.  We found that retaining the first $5$ photon levels was sufficient to attain convergence.


\section{Unitary dynamics}\label{sec:unitary}

In this section we  focus  on the unitary dynamics described by the Hamiltonian $H_\text{s}(t)$ defined in Eq.(\ref{eq:H0t}). It will be useful to define the set of Bell states of the two qubit system: $\ket{\Phi_\pm} \equiv \frac{1}{\sqrt{2}}(\ket{\uparrow\uparrow}\pm\ket{\downarrow\downarrow})$ and $\ket{\Psi_\pm} \equiv \frac{1}{\sqrt{2}}(\ket{\uparrow\downarrow}\pm\ket{\downarrow\uparrow})$, where $\sigma_z\ket{\uparrow}=\ket{\uparrow}$, and $\sigma_z \ket{\downarrow}=-\ket{\downarrow}$.  The Bell states are maximally entangled and form a basis for the two qubit Hilbert space.  

Figure \ref{fig:spectrum}(a) shows the energy spectrum of $H_\text{s}$ parametrized as a function of $\epsilon$, the driving variable. It can be shown that the avoided  crossings at $\epsilon=\pm \Omega$ have  energy gaps of first order in $g_i$  while for all other integer and half integer values of $\epsilon/\Omega$ the avoided crossing gaps  are of  second or higher order in $g_i$. Away from all avoided crossings, the energies and eigenstates of the system satisfy

\begin{align}
    H_\text{s}(\epsilon)\ket{N \uparrow\uparrow}\  &\approx\  (N\Omega+\epsilon)\ket{N \uparrow\uparrow}\\
    H_\text{s}(\epsilon)\ket{N \Psi_\pm}\ &\approx\ \ \ \ \ \ \ \ N\Omega\ket{N \Psi_\pm}\\
    H_\text{s}(\epsilon)\ket{N \downarrow\downarrow}\  &\approx\  (N\Omega-\epsilon)\ket{N \downarrow\downarrow},
\end{align}
with $\ket{N}$ the state of the resonator with $N$ photons. In addition, it can be readily seen from Eq.(\ref{eq:H0t}) that for $g_1=g_2$
($\delta_g=0$), the singlet states $\ket{N\Psi_-}$ are exact eigenstates of  $H_\text{s}(\epsilon)$  with  energy $N\Omega$. Since they  are also eigenstates of the driving operator ($\propto (\sigma_z^{(1)}+ \sigma_z^{(2)}) $), transitions involving the states $\ket{N\Psi_-}$ are forbidden for $\delta_g=0$. In fact, defining $g\equiv (g_1+g_2)/2$ we can
write
\begin{equation}
    \sum_i g_i \sigma_x^{(i)} = g(\sigma_x^{(1)}+\sigma_x^{(2)}) + \frac{\delta_g}{2}(\sigma_x^{(1)}-\sigma_x^{(2)}),\label{eq:g_divided}
\end{equation}
and consider the  second  term  in the last equation     as a  perturbation that induces transitions between $\ket{N\Psi_-}$ and $\ket{(N\pm 1)\Phi_+}$, since we suppose that $\delta_g \ll g$. This can be seen from the fact that
\begin{equation}
    \frac{\delta_g}{2}(\sigma_x^{(1)}-\sigma_x^{(2)}) = \delta_g\Big(\ket{\Psi_-}\bra{\Phi_+} +   \ket{\Phi_+}\bra{\Psi_-}\Big)\label{eq:delta_g},
\end{equation}
and that  the $(a+a^\dagger)$ operator  either adds or removes a photon from the resonator state which it acts on.

\begin{widetext}

\begin{figure}[h]
    \centering
    \includegraphics[width=\linewidth]{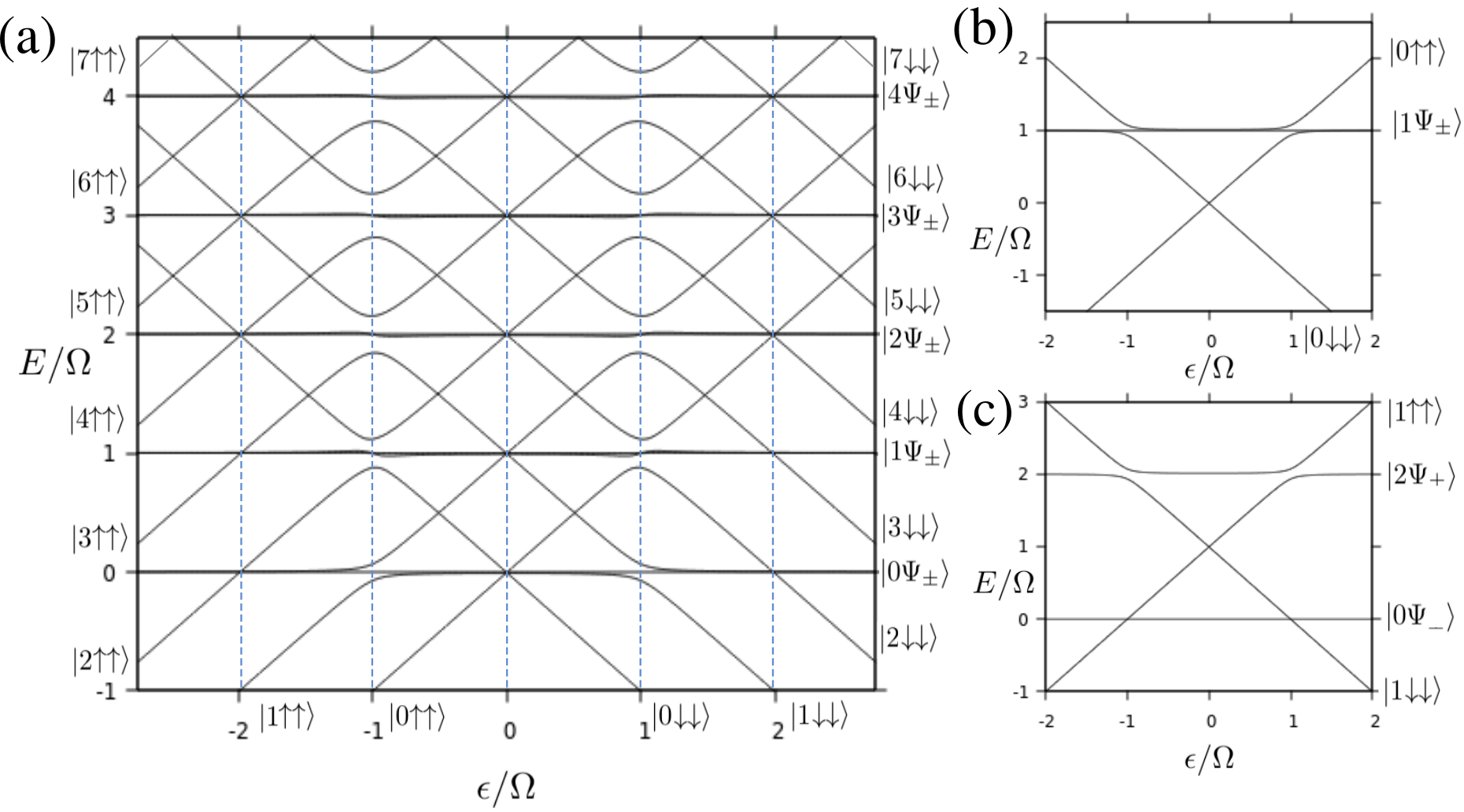}
    \caption {(a) Energy spectrum of $H_\text{s}$ (Eq.(\ref{eq:H0t})) parametrized  a function of  $\epsilon/\Omega$, for $g_1=0.05\Omega$ and $g_2=0.0485\Omega$. 
    Avoided crossings are located at integer and half-integer values of $\epsilon/\Omega$. 
    Those at $\epsilon=\pm \Omega$  are of first order in $g_i$ while all others  are of second and higher order in $g_i$.
    Integer values of $\epsilon/\Omega$ are marked with dotted lines. 
    See text for more details. Energy spectra of the effective Hamiltonians (Eq.(\ref{eq:H_long}) and  Eq.(\ref{eq:H_trans})) derived to study the transition (b) $\ket{0\uparrow\uparrow}\rightarrow\ket{1\Psi_-}$ and  the transition (c) $\ket{0\Psi_-}\rightarrow$(other states), respectively. In all cases the asymptotic eigenstates away from avoided crossings are shown.}
    \label{fig:spectrum}
\end{figure}

\begin{figure}[h]
    \centering
    \includegraphics[width=\linewidth]{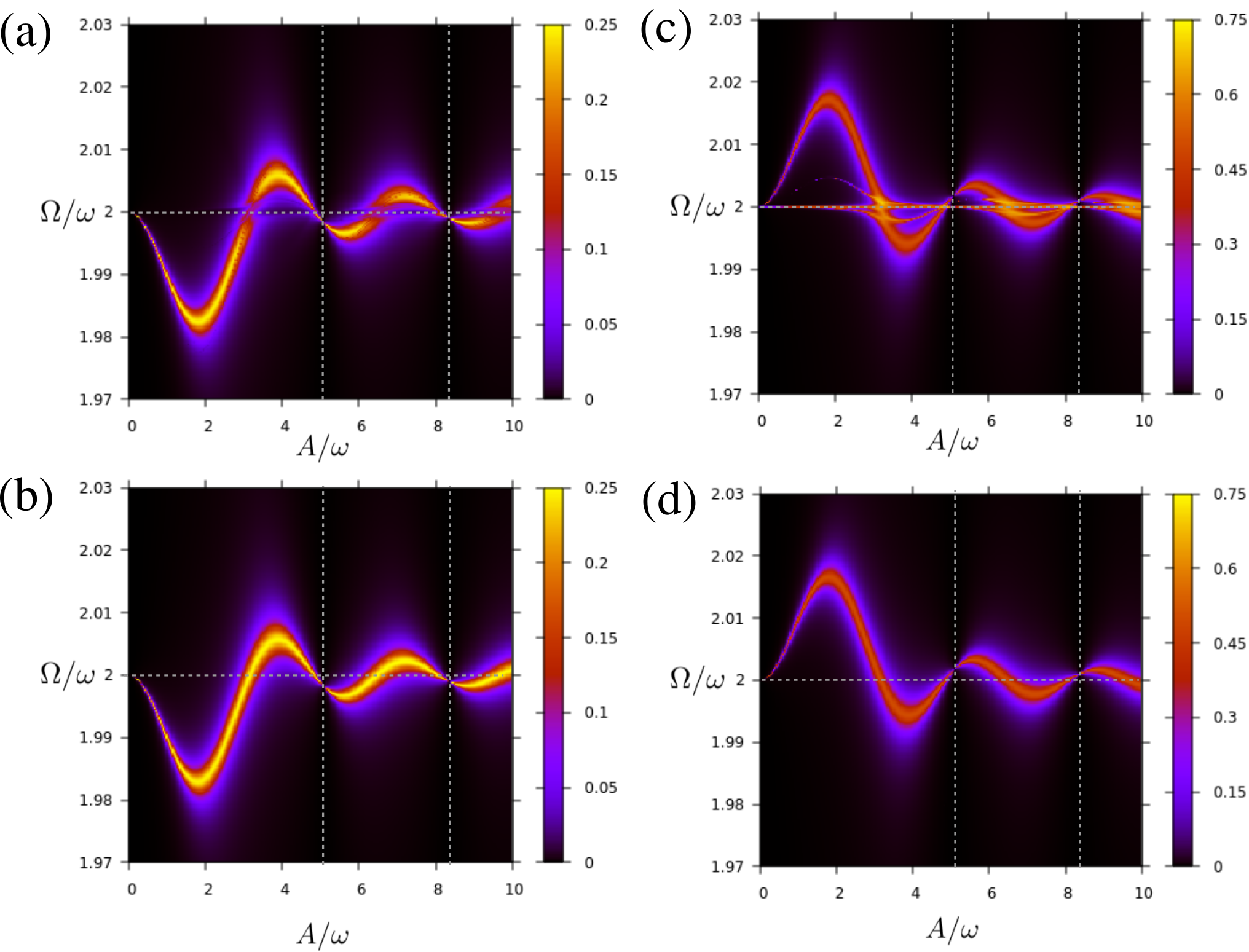}
    \caption{Upper panel: Unitary time-averaged transition probabilities calculated numerically  with the Hamiltonian Eq.(\ref{eq:H0t}) for the transitions  (a) $\ket{0\uparrow \uparrow}\rightarrow\ket{1\Psi_-}$  and (c) $\ket{0\Psi_-}\rightarrow$ (other states) , as a function of $A/\omega$ and $\Omega/\omega$. The patterns show resonances at $\Omega/\omega=n\in \mathbb{N}$ with width of order $\delta_g$ modulated by the $n^\text{th}$ Bessel function $J_n(A/\omega)$.
   Lower panel: Transition probabilities for (b) $\ket{0\uparrow \uparrow}\rightarrow\ket{1\Psi_-}$  and (d) $\ket{0\Psi_-}\rightarrow$ (other states) , calculated using the effective two-qubit Hamiltonians Eq.(\ref{eq:H_long}) and Eq.(\ref{eq:H_trans}), respectively.  The line $\Omega/\omega=2$ and the lines for which $A/\omega$ equals the first two zeros of the $2^\text{nd}$ Bessel function are indicated with grey dotted lines.
    In all cases the qubit-resonator coupling strengths used are $g_1=0.1\omega$ and $g_2=0.097\omega$. }\label{fig:unit}
\end{figure}

\end{widetext}

In what follows we consider explicitly transitions involving $\ket{0\Psi_-}$ and $\ket{1\Psi_-}$ induced by the simultaneous effect of the time dependent driving  $\epsilon(t)$, Eq.(\ref{eq:driving}), and the  coupling asymmetry parameter $\delta_g$. These transitions will be relevant for the Bell state generation mechanism once dissipation is included.

\subsection{Transitions to $\ket{1\Psi_-}$}\label{S3A}

We begin by studying  the transition probability of $\ket{0\uparrow\uparrow}\rightarrow \ket{1\Psi_-}$ induced by the periodic driving.
We  are interested in this transition because, as we will discuss in Sec.\ref{sec:bell},   $\ket{1\Psi_-}$ will decay into $\ket{0\Psi_-}$ after including dissipation, which is a maximally entangled state that is stable against photon loss.

Figure \ref{fig:unit}(a) shows the time averaged transition probability  for $\ket{0\uparrow\uparrow}\rightarrow \ket{1\Psi_-}$, as a function of $A/\omega$ and $\Omega/\omega$, near a resonance centered around   $\Omega/\omega=2$. Similar resonances of width proportional to $\delta_g$ are observed for all integer values of $\Omega/\omega$. 

Numerically  we find  that for these resonances, and for  $\ket{0\uparrow\uparrow}$ as the initial state, the most populated states are in
the subspace $S_1$ spanned by
$\{\ket{0\Phi_\pm},\ket{1\Psi_\pm}\}$. Projecting the Hamiltonian of Eq.(\ref{eq:H0t}) into $S_1$, the state of the resonator becomes uniquely determined by the state of the qubits (within $S_1$). Thus one can write the terms involving resonator operators in  Eq.(\ref{eq:H0t}) in terms of qubit operators:

\begin{align}
    a^\dagger a |_{S_1}   =& \ket{1\Psi_+}\bra{1\Psi_+} + \ket{1\Psi_-}\bra{1\Psi_-} \nonumber\\
    =& \frac{1}{2}(1-\sigma_z^{(1)} \sigma_z^{(2)}),
\end{align}
and
\begin{align}
    \frac{1}{2}(a+a^\dagger)(\sigma_x^{(1)}\pm\sigma_x^{(2)})|_{S_1}=& \ket{0\Phi_\mp}\bra{1\Psi_\pm} + \ket{1\Psi_\pm}\bra{0\Phi_\mp} \nonumber\\
    =& \frac{1}{2}(\sigma_x^{(1)}\pm\sigma_x^{(2)}),
\end{align}
where now $\sigma^{(i)}_j$ are understood as $4\times 4$ matrices, and $|_{S_1}$ indicates projection into $S_1$.
Replacing the above expressions into Eq.(\ref{eq:H0t}), one arrives at the  effective time dependent Hamiltonian,  valid for studying the transition  $\ket{0\uparrow\uparrow}\rightarrow\ket{1\Psi_-}$:

\begin{equation}
    H_{l}(t) = \sum_i\Big(\frac{\epsilon (t)}{2}\sigma_z^{(i)} + g_i \sigma_x^{(i)}\Big) - \frac{\Omega}{2}\sigma_z^{(1)}\sigma_z^{(2)}+\frac{\Omega}{2}.\label{eq:H_long}
\end{equation}
 

Notice that Eq.(\ref{eq:H_long}) is the Hamiltonian of two driven qubits  coupled longitudinally studied in Ref.\onlinecite{Gramajo_2017}. 
 In the present case  $\Omega$ plays the role of the interaction strength between the qubits and  $2g_i$ the role of the intrinsic qubit gaps. 
Figure \ref{fig:spectrum}(b) shows the energy spectrum of the effective Hamiltonian, Eq.(\ref{eq:H_long}), parametrized as a function of  $\epsilon$.

 In Fig. \ref{fig:unit}(b) the time averaged transition probability  $\ket{0\uparrow\uparrow}\rightarrow\ket{1\Psi_-}$  computed numerically using  Eq.(\ref{eq:H_long}) is displayed. The  agreement with Fig. \ref{fig:unit}(a), obtained from  the full Hamiltonian of Eq.(\ref{eq:H0t}), is excellent. 
 
\begin{figure}[h]
    \centering
    \includegraphics[width=\linewidth]{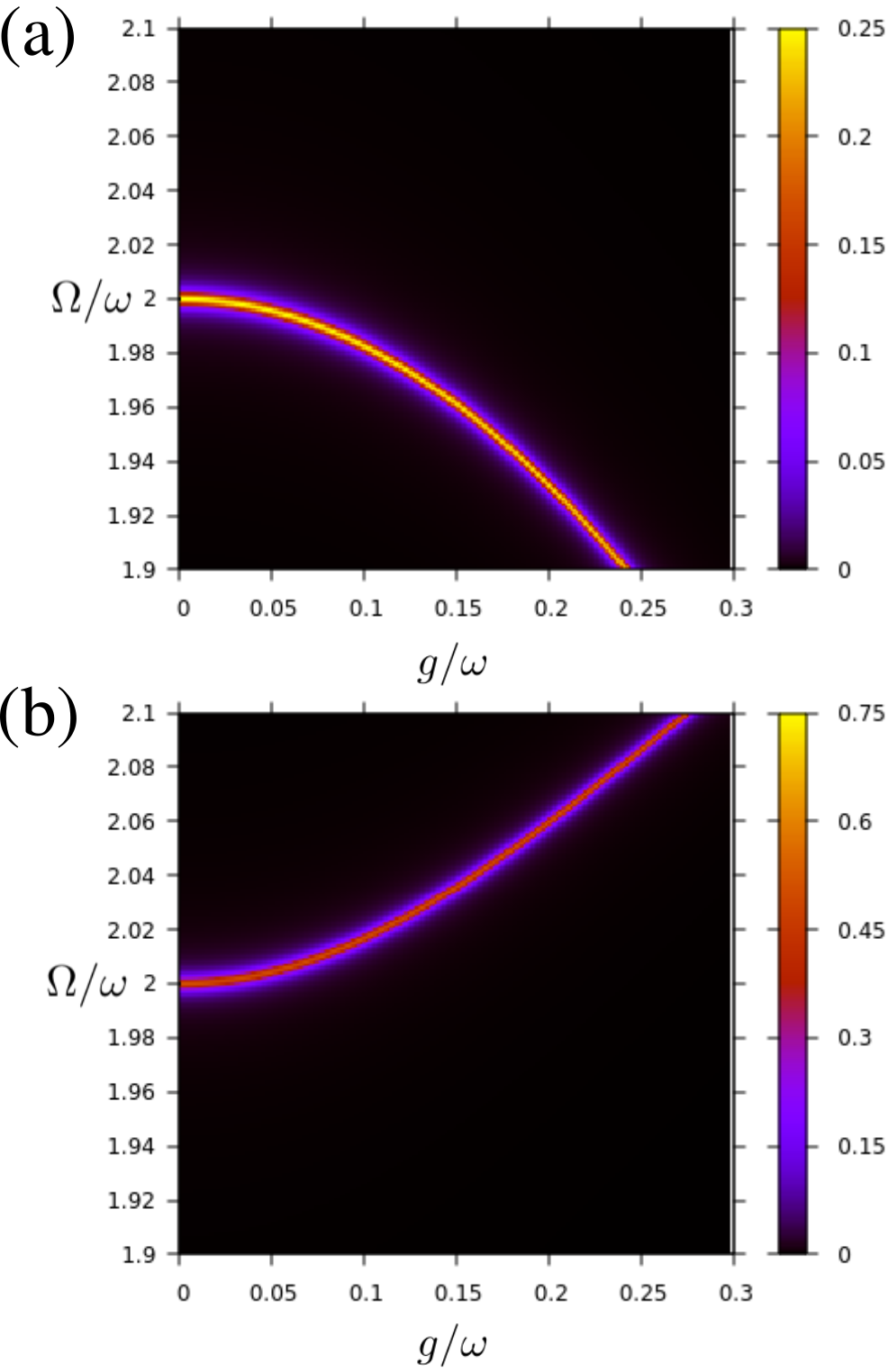}
    \caption{Unitary time averaged transition probabilities for the transitions (a) $\ket{0\uparrow\uparrow}\rightarrow\ket{1\Psi_-}$ and (b) $\ket{0\Psi_-}\rightarrow$(other states) calculated using the effective Hamiltonians Eq.(\ref{eq:H_long}) and Eq.(\ref{eq:H_trans}) respectively, as a function of $g/\omega$ and $\Omega/\omega$, for $\delta_g = 0.003\omega$ and $A=\Omega$. For small $g$, an approximately quadratic dependence of the resonance displacement as a function of $g$ is observed, with opposite curvature for both transitions.}
    \label{fig:g}
\end{figure}

For $\delta_g^2 \ll A\omega$ and neglecting the effect of $g$ on the energy spectrum, the LZS resonance condition for the transition between the two levels $\ket{0\uparrow\uparrow}$ and $\ket{1\Psi_-}$ is  $\Omega/\omega=n$ for some integer $n$, where in this case $\Omega$ is playing the same  role of the dc detuning in standard LZS interferometry, as it sets the average energy difference between the two involved levels. 
Thus, considering only these two levels which are separated by  a gap of the order of $\delta_g$, resonance patterns centered around $\Omega/\omega=n$ and of width $\propto \delta_g J_n(A/\omega)$ (with $J_n$ the $n^\text{th}$ Bessel function)
  are expected \cite{Ashhab_2007} and indeed  observed in Fig. \ref{fig:unit}(b). However, the curvature of the resonances is an effect that
  we found numerically to be of order $g_1 g_2 /\Omega$, as shown in Fig. \ref{fig:g}(a). This effect is not captured under the assumption of neglecting $g$. 

\subsection {Transitions out of $\ket{0\Psi_-}$}

We now shift our attention to the time averaged probabilities for the $\ket{0\Psi_-}\rightarrow \text{(other states)}$ transition, which is defined as the sum of all transition probabilities from the initial state $\ket{0\Psi_-}$ to any other states different from it. We are interested in this transition because it is an indicator of the stability of $\ket{0\Psi_-}$ against unitary transitions induced by the driving that  might take the system out of this state. Figure \ref{fig:unit}(c) shows numerical results for the corresponding transition probabilities using the full Hamiltonian Eq.(\ref{eq:H0t}). We have found  that the only significantly populated states are in the subspace  $S_2$ spanned by $\{\ket{0\Psi_\pm},\ket{1\Phi_\pm},\ket{2\Psi_+}\}$.  
Thus in the present case, and unlike the analysis of Sec.\ref{S3A},  $5$ linearly independent states are in principle involved in the transition under study.

Numerically  we found that the best two-qubit effective model is obtained when neglecting the state $\ket{2\Psi_+}$ in the  calculation of the operator $(a+a^\dagger)|_{S_2}$, and the state $\ket{0\Psi_+}$ in the  calculation of  the operator $a^\dagger a|_{S_2}$. Doing this approximation, one obtains

\begin{align}
    a^\dagger a|_{S_2} \approx& \ket{1\Phi_+}\bra{1\Phi_+} + \ket{1\Phi_-}\bra{1\Phi_-} + 2\ket{2\Psi_+}\bra{2\Psi_+} \nonumber\\
    =& 1 + \frac{1}{2}(\sigma_x^{(1)}\sigma_x^{(2)}+\sigma_y^{(1)}\sigma_y^{(2)}),
\end{align}

\begin{align}
    \frac{1}{2}(a+a^\dagger)(\sigma_x^{(1)}\pm\sigma_x^{(2)})|_{S_2} \approx& \ket{0\Psi_\pm}\bra{1\Phi_\mp} + \ket{1\Phi_\mp}\bra{0\Psi_\pm} \nonumber\\
    =& \frac{1}{2}(\sigma_x^{(1)}\pm\sigma_x^{(2)}),
\end{align}
and we arrive at  the effective Hamiltonian, valid for studying the transition $\ket{0\Psi_-}\rightarrow$(other states):
\begin{align}
    H_\text{tr}(t) =& \sum_i\Big(\frac{\epsilon(t)}{2}\sigma_z^{(i)} + g_i \sigma_x^{(i)}\Big) + \nonumber\\ &  
    \frac{\Omega}{2}(\sigma_x^{(1)}\sigma_x^{(2)}+\sigma_y^{(1)}\sigma_y^{(2)}) + \Omega , \label{eq:H_trans}
\end{align}


which is the Hamiltonian of two transversally coupled and symmetrically driven qubits, with $\Omega$ playing the role of the interaction strength and $2g_i$ the role of the intrinsic qubit gaps \cite{Gramajo_2017}. Figure \ref{fig:spectrum}(c) shows its energy spectrum  parametrized as a function of $\epsilon$.
Numerical results for the transition probability   $\ket{0\Psi_-}\rightarrow\text{(other states)}$ computed from  Eq.(\ref{eq:H_trans}) are shown in 
Fig.\ref{fig:unit}(d). Except for the additional flatter resonances around integer $\Omega/\omega$, which correspond to resonances to states outside $S_2$, the  agreement with Fig.\ref{fig:unit}(c), obtained  using the full Hamiltonian Eq.(\ref{eq:H0t}) is remarkable. 

The LZS resonance  condition for both transitions, $\ket{0\Psi_-} \rightarrow \ket{1\uparrow\uparrow}$ and  $\ket{0\Psi_-}\rightarrow \ket{1\downarrow\downarrow}$ respectively, is again $\Omega/\omega=n$ for some integer $n$. Resonances around these values of $\Omega/\omega$ of width $\propto \delta_g J_n(A/\omega)$ are expected and  observed in Fig. \ref{fig:unit}(d), in analogy to the results of Sec.\ref{S3A}.
The effect of $g$ on the curvature of the resonances out of $\ket{0\Psi_-}$ is again of order $g_1 g_2 /\Omega$ but of opposite sign to that of the transitions to $\ket{1\Psi_-}$, as can be seen in Fig. \ref{fig:g}(b). This implies that there are values of the  driving parameters, $A$ and $\omega$, for which the transitions to $\ket{1\Psi_-}$ are stimulated but those  out of $\ket{0\Psi_-}$ are not. This is the key point that will be made use of to  generate Bell states once dissipation is included, as is explained below.










\section{Dissipation induced Bell state generation\label{sec:bell}}

 In the previous section we concluded  that starting from the initial state $\ket{0\uparrow\uparrow}$, there are regions  in the $A/\omega-\Omega/\omega$ plane where a unitary resonance to $\ket{1\Psi_-}$ is stimulated but no unitary resonance involving $\ket{0\Psi_-}$ is so. When the driving amplitude A and frequency $\omega$ are chosen as to  select one of these points, the process
\begin{equation}
    \ket{0\uparrow\uparrow}\stackrel{H_\text{s}(t)}{\rightarrow}\ket{1\Psi_-}\stackrel{\text{PL}}{\rightarrow}\ket{0\Psi_-}\label{eq:process}
\end{equation}
can occur once dissipation is included, where the first transition is unitary and induced by the driven Hamiltonian $H_{s}(t)$, and the second one is the loss of a photon to the environment.
The same process can also take place when starting from $\ket{0\downarrow\downarrow}$ or $\ket{0\Psi_+}$, since they show similar unitary resonances for $\ket{0\downarrow\downarrow}\rightarrow\ket{1\Psi_-}$ and $\ket{0\Psi_+}\rightarrow\ket{1\Psi_-}$.

Since the open system dynamics is linear due to the assumed weak coupling to the environment, one can understand the evolution of the system's density matrix as the independent evolution of its ensemble members
which, for $T_\text{b} \ll \Omega$,  will eventually reach one of the states with zero photons. 
If the reached states are different from $\ket{0\Psi_-}$,  they will go through the process defined in Eq.(\ref{eq:process})  ending up at least partially in $\ket{0\Psi_-}$. As  all transitions involving $\ket{0\Psi_-}$ are out of resonance,  gradually the population of the system's density matrix  accumulates in this state.

\begin{figure}[h]
    \centering
    \includegraphics[width=\linewidth]{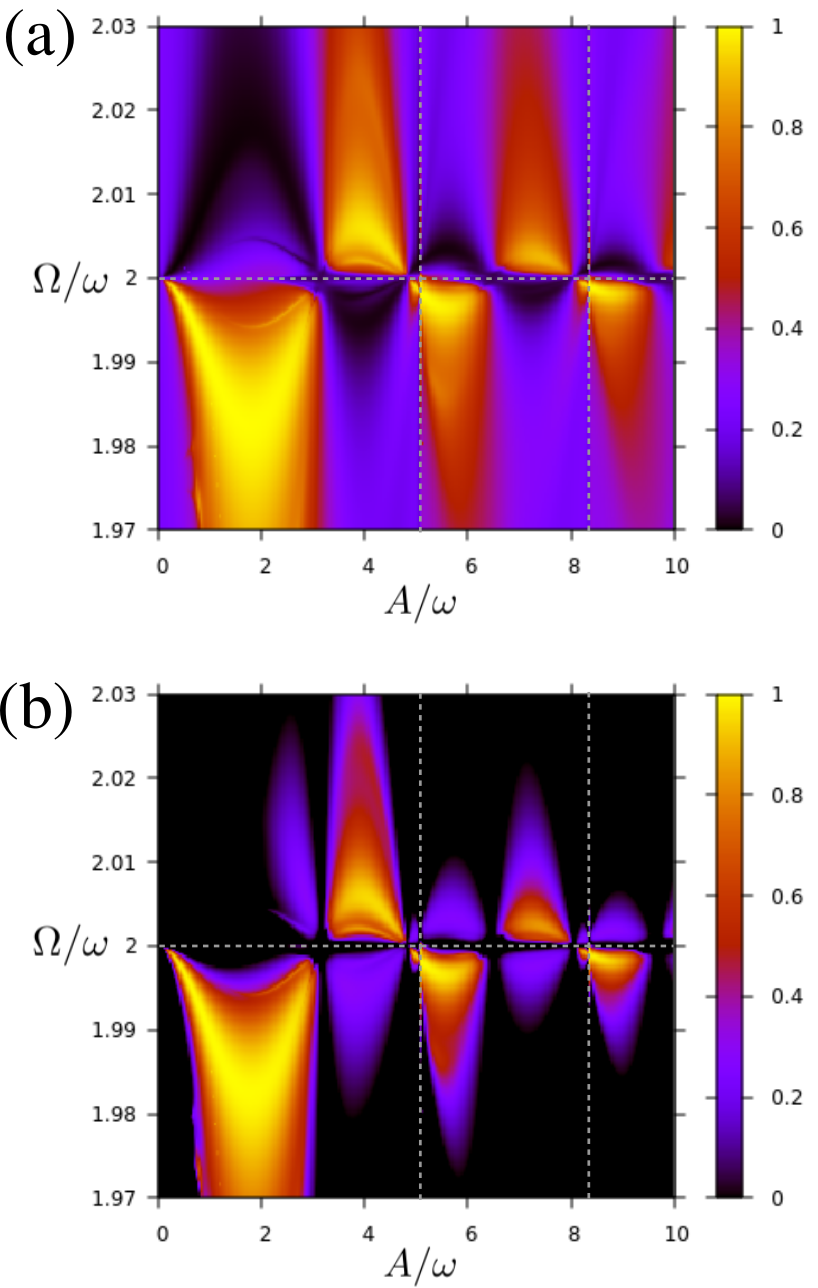}
    \caption{Time averaged  (a) population of $\ket{0\Psi_-}$ and  (b) concurrence of the steady state of the driven dissipative system of two qubits coupled to a resonator, obtained after solving numerically the FBM master equation associated to the Hamiltonian defined in Eq.(\ref{eq:H_3parts}). Asymmetrical resonance patterns
    are observed  at integer values of $\Omega/\omega$. The coupling strengths are the same as in Fig. \ref{fig:unit}, with bath parameters $\kappa\lambda^2=0.0001$ and $T_\text{b}=0.001\omega$. The line $\Omega/\omega=2$ and the lines for which $A/\omega$ equals the first two zeros of the $2^\text{nd}$ Bessel function are indicated with grey dotted lines.}
    \label{fig:diss}
\end{figure}

Figure \ref{fig:diss}(a) shows numerical results for the time averaged population of $\ket{0\Psi_-}$ in the stationary state as a function of  $A/\omega$ and $\Omega/\omega$ obtained after solving numerically the FBM master equation for the system's density matrix $\rho$ (see Appendix B for details). The stationary state is found to be unique and  $T$-periodic in time. The behavior of the stationary population closely follows what is predicted by the previous argument: population maxima very close to $1$ are observed for points in the parameter space along the unitary resonance patterns of $\ket{0\uparrow\uparrow}\rightarrow\ket{1\Psi_-}$ and population minima, very close to $0$, are obtained for points along the unitary $\ket{0\Psi_-}\rightarrow\text{(other states)}$ resonances.

To quantify the degree of entanglement of the qubits, we use the concurrence as a measure \cite{Wootters_1998},

\begin{equation}
    \mathcal{C}[\rho_\text{q}] = \max\{0,r_3-r_2-r_1-r_0\},\label{eq:conc}
\end{equation}

defined in terms of the qubits' density matrix $\rho_\text{q}=\text{Tr}_\text{r}(\rho)$, where the trace operation is over the states of the resonator and  $r_i$ are the real-valued eigenvalues of

\begin{equation}
    \sqrt{\sqrt{\rho_\text{q}}\sigma_y^{(1)}\sigma_y^{(2)}\rho_\text{q}^*\sigma_y^{(1)}\sigma_y^{(2)}\sqrt{\rho_\text{q}}},
\end{equation}

sorted in ascending order, where  $\rho_\text{q}^*$ is the complex conjugate of $\rho_\text{q}$ and  the conjugation must be done in a separable basis.
The concurrence takes a value of $0$ for a separable state, a value of $1$ for a maximally entangled state, and values in between for partially entangled states.

Figure \ref{fig:diss}(b) shows the time averaged concurrence of the stationary state. It is observed that the maxima of the  concurrence that are close to $1$ are achieved only when the state $\ket{0\Psi_-}$ is populated, indicating that this is the only maximally entangled state that is generated. It is also noteworthy that for driving parameters lying outside the mentioned resonances, which   constitute the vast majority of the points in the $A/\omega-\Omega/\omega$ plane (including the case of no driving at all $A=0$), the stationary state is either separable or almost separable.

As we have already mentioned, to attain  $\ket{0\Psi_-}\bra{0\Psi_-}$ as the stationary state of the system, a necessary condition is to find points in the plane  $A/\omega-\Omega/\omega$  where the resonance conditions for the transitions to $\ket{1\Psi_-}$ and   out of $\ket{0\Psi_-}$  do not overlap. 
This can only be fulfilled if  the maximum resonance deviation from the condition $\Omega/\omega=n$ (of order $g_1g_2/\Omega$) is much greater than the resonance width (of order $\delta_g$), and this the source of the requirement $\delta_g \ll g_1 g_2/\Omega$ . 
The optimal entanglement generation (maximal area and intensity of concurrence patterns)  is achieved for amplitudes $A=\Omega$ and frequencies $\omega$ such that $\Omega$ is below an integer multiple of $\omega$ by a frequency of the order of $g_1 g_2 /\Omega$, i.e. $A=\Omega=n\omega - \mathcal{O}(g_1 g_2 /\Omega)$.

\begin{figure}[h]
    \centering
    \includegraphics[width=\linewidth]{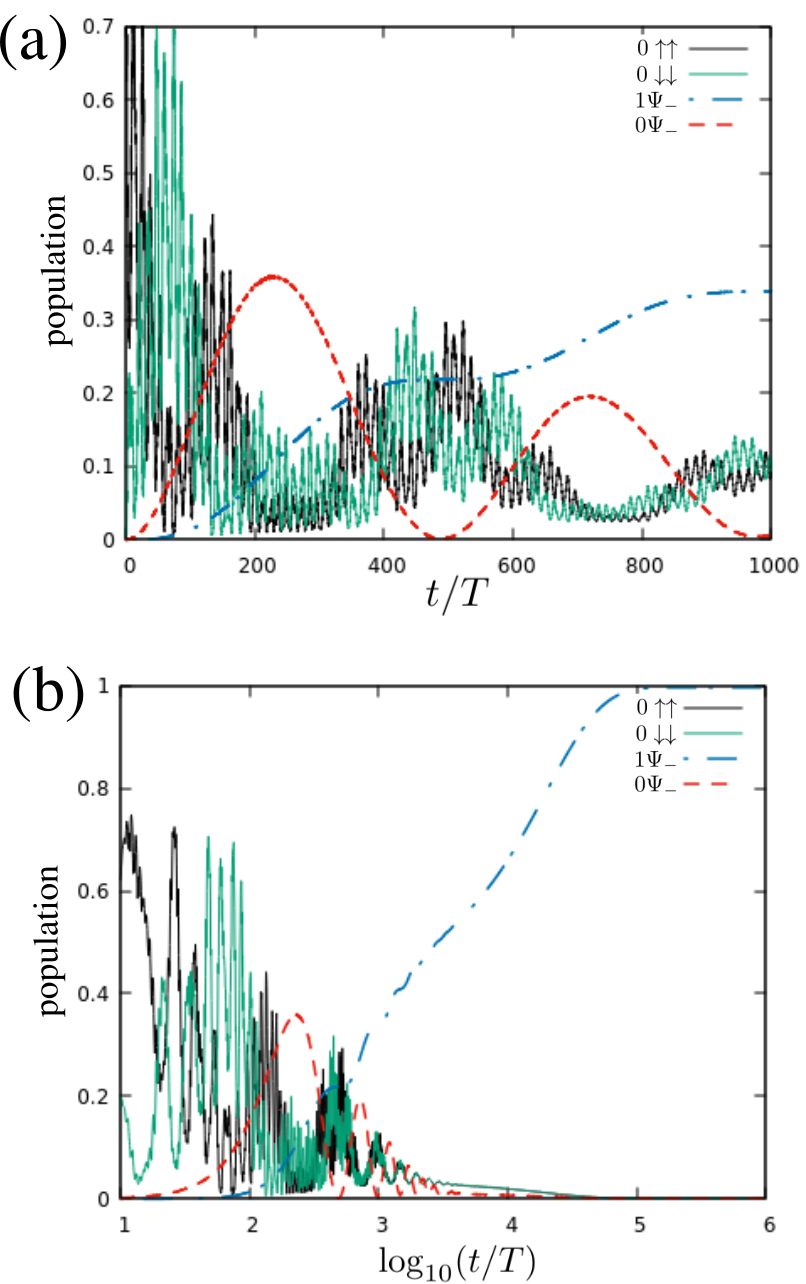}
    \caption{Temporal dependence of the system's density matrix populations, starting from the initial state $\ket{0\uparrow\uparrow}\bra{0\uparrow\uparrow}$, for up to (a) $1000$ (linear scale) and (b) $10^6$ (log scale) driving periods, dissipation included. The driving parameters are $A=\Omega=1.983\omega$, and the rest of the parameters are the same as in Fig.\ref{fig:diss}. 
    The populations of $\ket{0 \uparrow \uparrow}$, $\ket{0 \downarrow \downarrow}$, $\ket{0 \Psi_-}$ and $\ket{1 \Psi_-}$ are shown. It is seen that for short times a resonance between $\ket{0 \uparrow \uparrow}$, $\ket{0 \downarrow \downarrow}$ and $\ket{1 \Psi_-}$ is stimulated. For later times and via photon loss  population gradually accumulates in $\ket{0 \Psi_-}$.}
    \label{fig:temp}
\end{figure}

Finally, Fig.\ref{fig:temp} shows the temporal dynamics of relevant populations of the system's density matrix  at a point of high entanglement generation $A=\Omega=1.983\omega$, for the system starting in the state $\rho_0 = \ket{0\uparrow\uparrow}\bra{0\uparrow\uparrow}$. It is seen that even though the dynamics of the populations is complicated, a clear  resonance to the state $\ket{1\Psi_-}$ is stimulated in a timescale of the order of $\delta_g^{-1}$, which gradually decays into $\ket{0\Psi_-}$ via photon loss to the environment. Population accumulates in this state, and 
given enough time the system ends up essentially at $\rho_\infty = \ket{0\Psi_-}\bra{0\Psi_-}$.

We thus conclude that by applying a driving with appropriate amplitude and frequency it is possible to populate the maximally entangled state $\ket{0\Psi_-}$ independently of the initial state of the system. 

\section{Conclusions}\label{sec:concl}

In this work we have presented an entanglement generation protocol for a system of two qubits coupled through a resonator.  A maximally entangled steady state is achieved when a symmetric ac driving  is applied over both qubits with appropriate amplitude and frequency.

The areas of entanglement generation in the $A/\omega$-$\Omega/\omega$ plane 
 are associated to resonant unitary transitions into $\ket{1\Psi_-}$ and out of $\ket{0\Psi_-}$. If the driving is such that transitions to $\ket{1\Psi_-}$ are in resonance, but transitions out of $\ket{0\Psi_-}$ are not, once dissipation is included the system will accumulate population in $\ket{0\Psi_-}$ via photon loss from $\ket{1\Psi_-}$.
 All the relevant features of the unitary resonance patterns were  described  in terms of effective  Hamiltonians for two driven qubits.

The optimal steady state  entanglement generation (in terms of  maximal area and intensity) is attained for driving amplitudes $A=\Omega$ and frequencies such that $\Omega/\omega$ is slightly below an integer number.

The proposed  protocol shows advantages with respect to other methods for entanglement generation. 
Due to dissipation, the steady state is independent of the initial state of the system \cite{kraus_2008,verstraete_2009} and, once the entangled state is obtained, it is protected from environmental effects for as long as the driving is applied. 
Moreover, the present proposal allows entangling  distant and strongly driven qubits which are, for example, a microwave waveguide apart. Therefore, it is expected that our scheme could add a robust means to realize entanglement protocols in  
setups extensively used
nowadays in cQED  \cite{Blais_2020, Blais_2021}.

\section*{Acknowlegments}
We acknowledge support from CNEA, CONICET (PIP11220150100756), ANPCyT (PICT2016-0791 and PICT2019-0654) and  UNCuyo (06/C591).

\appendix 

\section{Model of the environment\label{app:env}}

We model the bath and its interaction with the system using the Caldeira-Leggett model \cite{breuer_petruccione_2006}:
 
\begin{align}
    H_\text{b} = &\int_0^\infty d\omega \omega b_\omega^\dagger b_\omega\\
    H_\text{sb} =& \lambda A\int_0^\infty d\omega \sqrt{\mathcal{J}(\omega)}(b_\omega+b_\omega^\dagger)+H_\text{rn}\\
    H_\text{rn}=&\lambda^2 A^2\int_0^\infty d\omega \frac{\mathcal{J}(\omega)}{\omega},
\end{align}

where $b_\omega$ and $b_\omega^\dagger$ are the creation and destruction operators of the harmonic oscillator continuum, $A$ is a unitless system operator, $\lambda$ is a coupling strength, and $H_\text{rn}$ is a renormalization term to cancel all Lamb shifts induced on the system by the thermal bath.
 
For the system under study we chose $A=a+a^\dagger$ and $\mathcal{J}(\omega)=\kappa \omega$, with $\kappa$ a constant with units of energy$^{-2}$. Usually, it is necessary to put a cutoff frequency in the bath spectral density, but for our purposes, and since we have already cancelled out the Lamb shifts,  it is permissible to take this cutoff frequency as infinity (greater than all energy scales of the problem) as we have implicitly done by the choice of $\mathcal{J}(\omega)$.

\section{Floquet-Born-Markov master equation\label{app:fbm}}

Floquet theory is widely used to study time periodic unitary quantum systems \cite{Shirley_1965}. It shows that for a quantum system with a $T$-periodic Hamiltonian $H(t)$, all solutions are a linear combination of a single basis of states of the form $e^{-i\epsilon_\alpha t}\ket{u_\alpha(t)}$.  Here, $\ket{u_\alpha(t)}$ is $T$-periodic and is called a Floquet state, and $\epsilon_\alpha$ is called its corresponding quasienergy. 
To  find  the Floquet states and their quasienergies it is customary to  diagonalize the evolution operator $U(t,t_0)$ over a period of the driving, since the Floquet states satisfy the eigenvalue equation:

\begin{equation}
    U(t+T,t)\ket{u_\alpha(t)}=e^{-i\epsilon_\alpha T}\ket{u_\alpha(t)}.
\end{equation}

The Floquet-Born-Markov master equation \cite{kohler_1998,Ferron_2012,Blattmann_2015} allows modelling dissipative processes in periodically driven systems for sufficiently weak coupling to the environment. It is a linear Markovian differential equation for the time evolution of the system's density matrix: 

\begin{equation}
	\partial_t \rho_{\alpha \beta}(t) = -i(\epsilon_\alpha-\epsilon_\beta)\rho_{\alpha\beta}(t)+\sum_{\alpha'\beta'}L_{\alpha\beta\alpha'\beta'}(t) \rho_{\alpha' \beta'}(t)\label{eq:fbm_full},
\end{equation}

where $\rho_{\alpha\beta}(t)= \bra{u_\alpha(t)}\rho(t)\ket{u_\beta(t)}$ are the components of the system's density matrix in a Floquet basis.
The first term in (\ref{eq:fbm_full}) corresponds to the unitary evolution of the system, while the second one  takes into account  dissipative effects. The transition rates $L_{\alpha \beta \alpha' \beta'}(t)$  are $T$-periodic and can be  Fourier expanded as
\begin{equation}
	L_{\alpha \beta \alpha' \beta'}(t) = \sum_{q} L^{q}_{\alpha\beta \alpha' \beta'}e^{-iq\omega t},\label{eq:L_t}
\end{equation}
with $q\in\mathbb{Z}$, and where the coefficients  $L^{q}_{\alpha\beta \alpha' \beta'}$ are given by
\begin{align}
	&L^{q}_{\alpha\beta \alpha' \beta'}= \lambda^2\sum_{k} \Big( g_{\alpha\alpha'}^k A_{\alpha\alpha'}^k A_{\beta'\beta}^{-k-q}+g_{\beta\beta'}^k A_{\alpha\alpha'}^{k-q}A_{\beta'\beta}^{-k}\nonumber\\& - \sum_{\eta}(\delta_{\beta\beta'}g_{\eta\alpha'}^k A_{\alpha\eta}^{-k-q}A_{\eta\alpha'}^k+\delta_{\alpha\alpha'}g_{\eta\beta'}^k A_{\beta'\eta}^{-k}A_{\eta\beta}^{k-q}) \Big).
\end{align}

In the last expression, the index $\eta$ runs over the indices of the Floquet basis, $\delta_{\alpha\beta}$ is the Kronecker delta and we defined
\begin{equation}
	A_{\alpha\beta}^q = \sum_{k}\bra{u_\alpha^k}A\ket{u_\beta^{q+k}}\label{eq:A_ab}
\end{equation}
and
\begin{equation}
	g_{\alpha\beta}^k= g(\epsilon_\alpha-\epsilon_\beta+k\omega).\label{eq:g_ab}
\end{equation}
In Eq.(\ref{eq:A_ab}), $\ket{u_\alpha^k}$ is the $k$th Fourier component of the Floquet state $\ket{u_\alpha(t)}$. In Eq. (\ref{eq:g_ab}), $g(\omega')$ is the Fourier transformed correlation function of the thermal bath, which can be expressed in terms of its spectral density and the Bose occupation number $n_{T_\text{b}}(\omega') = 1/(e^{\omega'/T_\text{b}}-1)$ as
\begin{equation}
g(\omega')=
\begin{cases}
\mathcal{J}(\omega')n_{T_\text{b}}(\omega')\ \ \ &\Leftarrow \omega'>0\\
-\mathcal{J}(-\omega')n_{T_\text{b}}(\omega') &\Leftarrow \omega'<0,
\end{cases}
\end{equation}

with the value  of $g$ at $\omega'=0$ obtained by taking the appropriate limit.

For  sufficiently weak coupling to the environment, such that the maximum rate of relaxation or decoherence is much smaller than the driving frequency,
a  (moderate) RWA is justified in the transition rates, Eq.(\ref{eq:L_t}). 
This sets  the terms with $q\neq 0$ effectively to zero, yielding  the simplified expression:  

\begin{equation}
	L_{\alpha\beta \alpha'\beta'} \approx R_{\alpha\beta\alpha'\beta\beta'} + R_{\beta\alpha\beta'\alpha'}^* - \sum_\eta (\delta_{\beta\beta'}R_{\eta \eta \alpha'\alpha}+\delta_{\alpha\alpha'}R_{\eta\eta\beta'\beta}^*),\label{eq:FBM_mRWA_wrates}
\end{equation}

in terms of the quantities

\begin{equation}
	R_{\alpha\beta\alpha'\beta'} = \sum_k g_{\alpha\alpha'}^k A_{\alpha\alpha'}^k A_{\beta\beta'}^{-k}.\label{eq:FBM_rates}
\end{equation}

With this approximation, Eq.(\ref{eq:fbm_full}) no longer depends explicitly on time in the Floquet basis. 

For the cases studied in this work, it was found that the operator $\Lambda_{\alpha\beta\alpha'\beta'} = -i(\epsilon_\alpha-\epsilon_\beta)\delta_{\alpha\alpha'}\delta_{\beta\beta'}+L_{\alpha\beta\alpha'\beta'}$ can be numerically diagonalized in terms of left and right eigenvectors which are density matrices. That is,

\begin{equation}
	\Lambda \rho_\mu^\text{R} = \rho_\mu^\text{L} \Lambda = \zeta_\mu \rho_\mu,
\end{equation}

with $\zeta_\mu \in \mathbb{C}$ and $\frac{1}{N}\text{Tr}(\rho_\mu^\text{L}\rho_\nu^\text{R})=\delta_{\mu\nu}$, where $N$ is the dimension of the system's Hilbert space. Once these eigenvectors and eigenvalues are obtained, density matrices can be evolved readily by projecting on this eigensystem,

\begin{equation}
	\rho(t) = \sum_\mu c_\mu e^{\zeta_\mu (t-t_0)}\rho_\mu^\text{R},\ \ 
	c_\mu = \frac{1}{N}\text{Tr}(\rho_\mu^\text{L}\rho(t_0)).
\end{equation}
The real parts of $\zeta_\mu$ (which are always negative) are the decoherence and relaxation rates. The stationary state $\rho_\infty$, which for all cases studied in this work  can be found and is unique, is defined (in the Floquet basis) as the state $\rho^\text{R}_\mu$ with $\zeta_\mu=0$. It is constant in the Floquet basis and therefore $T$-periodic in the original system basis.

To calculate time averaged functions $f(\rho)$ of the system's density matrix  in the stationary state, such as populations or concurrence, we make use of the periodicity of $\rho_\infty$ and numerically integrate

\begin{equation}
    \bar{f} = \frac{1}{T}\int_0^T dt f(\rho_\infty(t)).
\end{equation}


\bibliography{references}

\end{document}